\documentclass{article}
\usepackage{graphicx} 
\usepackage{hyperref}
\usepackage{placeins}
\usepackage{enumitem}
\usepackage{graphicx}
\usepackage{subcaption}
\usepackage{caption}
\usepackage{float}
\usepackage{booktabs}
\usepackage{multirow}
\usepackage{xcolor}
\usepackage{makecell}
\usepackage{amssymb}  
\usepackage{amsfonts}
\usepackage{amsmath}
\usepackage{indentfirst}
\usepackage{authblk}
\newcommand{\corrauthor}{\textsuperscript{*}}

\title{First experimental measurements of biophotons from Astrocytes and Glioblastoma cell cultures}
\author[1]{L. De Paolis\corrauthor}
\author[1]{E. Pace\corrauthor}
\author[2]{C. Mazzanti\corrauthor}
\author[2]{M. Morelli}
\author[2]{F. Di Lorenzo}
\author[3,4]{L. Tonello}
\author[1]{C. Curceanu}
\author[1]{A. Clozza}
\author[5]{M. Grandi}
\author[6]{I. Davoli}
\author[7]{A. Gemignani}
\author[3]{P. Grigolini}
\author[1]{M. Benfatto\corrauthor}

\affil[1]{Laboratori Nazionali di Frascati, Istituto Nazionale di Fisica Nucleare, Via E. Fermi 40, 00044, Frascati, Italy.}
\affil[2]{Fondazione Pisana per la Scienza ONLUS, Via Ferruccio Giovannini, 13, 56017, S. Giuliano Terme, Pisa, Italy.}
\affil[3]{Center for Nonlinear Science, University of North Texas, P.O. Box 311427, Denton, TX 76203-1427, USA.}
\affil[4]{Gioya Higher Education Institution, E305, The Hub Workspace, Triq San Andrija, SGN1612, San Gwann, Malta.}
\affil[5]{Istituto La Torre, Via M. Ponzio 10, 10141 Torino, Italy;}
\affil[6]{Dipartimento di Fisica, Università di “Tor Vergata”, Via della Ricerca Scientifica, 00133, Roma, Italy.}
\affil[7]{Department of Surgical, Medical and Molecular Pathology, Critical and Care Medicine, University of Pisa, 56126, Pisa, Italy}

\date{2$^{nd}$ October 2025}
\begin{document}

\maketitle
\begingroup
\renewcommand\thefootnote{\textsuperscript{*}}
\footnotetext{Corresponding authors: \texttt{Luca.DePaolis@lnf.infn.it}, \texttt{Elisabetta.Pace@lnf.infn.it}, \texttt{C.Mazzanti@fpscience.it}, \texttt{Maurizio.Benfatto@lnf.infn.it}}
\endgroup

\begin{abstract}
    Biophotons are non-thermal and non-bioluminescent ultraweak photon emissions, first hypothesised by Gurwitsch in 1924 as a regulatory mechanism in cell division, and then experimentally observed in living organisms. Today, two main hypotheses explain their origin: stochastic decay of excited molecules and coherent electromagnetic fields produced in biochemical processes. Recent interest focuses on the role of biophotons in cellular communication and disease monitoring. This study presents the first campaign of biophoton emission measurements from cultured astrocytes and glioblastoma cells, conducted at Fondazione Pisana per la Scienza (FPS) using two ultra-sensitive setups developed by the collaboration at the National Laboratories of Frascati (LNF-INFN) and at the University of Rome II - Tor Vergata. The statistical analyses of the data collected revealed a clear separation between cellular signals and dark noise, confirming the high sensitivity of the apparatuses. The Diffusion Entropy Analysis (DEA) was applied to the data to uncover dynamic patterns, revealing anomalous diffusion and long-range memory effects potentially related to intercellular signalling and cellular communication. These findings support the hypothesis that biophoton emissions encode rich information beyond intensity, reflecting metabolic and pathological states. The differences that emerged from the application of Diffusion Entropy Analysis to the biophotonic signals of Astrocytes and Glioblastoma are highlighted and discussed in the paper. This work lays the foundation for future studies on neuronal cultures and proposes biophoton dynamics as a promising tool for non-invasive diagnostics and cellular communication research. 
\end{abstract}

\textbf{Keywords:} biophotons; complexity; data analysis; astrocyte; glioblastoma; cancer; tumour; diagnostic;

\section{Introduction}
Biophotons are ultraweak photon emissions observed in all living organisms, first hypothesized by Gurwitsch in 1924 \cite{Gurwitsch1}. Studying onion's root development, he showed that UV radiation from nearby roots enhanced mitosis without biochemical contact; a quartz (UV-transparent) barrier preserved the effect, while an opaque one suppressed it, suggesting a photonic regulatory mechanism called "morphogenetic field" \cite{Gurwitsch1, Gurwitsch2}.
In the following decades, emissions were detected in the UV \cite{Reiter, Rajewsky, Siebert, Audubert} and visible \cite{Colli1, Colli2} ranges from germinating seeds and other biological samples. In the 1980s, Popp coined the term \textit{biophoton} to describe this “mitogenic radiation” and began investigating its role \cite{Popp1}. Ultraweak photon emission (UPE), typically ~1–1000 photons/s/cm², is now recognized as a universal phenomenon in living systems, distinct from bioluminescence and not thermal radiation in nature \cite{Popp1, Quickenden, Wijk, Mayburov}.\par
Two main hypotheses address biophoton origins \cite{Popp1, Wijk}: (i) random radiative decay of metabolically excited molecules (e.g. oxidative reactions), and (ii) coherent EM fields generated by biochemical processes, possibly involving oxygen. Experimental data show that stress increases biophoton emission, supporting both models \cite{Slawinski1, Slawinski2}, which are not mutually exclusive.\par
Biophotons have recently attracted growing interest due to their potential role in cellular communication and regulation, with applications spanning toxicology, health monitoring, and cancer research \cite{Gallep, Tessaro, Popp2}. \par
Glioblastoma (GBM) is the most common and aggressive malignant primary brain tumour in adults. In the 2021 WHO classification, GBM is defined as an IDH-wildtype diffuse astrocytic tumour, CNS WHO grade 4, typically showing microvascular proliferation and/or necrosis \cite{Louis}. Standard of care remains maximal safe resection followed by radiotherapy with concomitant and adjuvant temozolomide (the Stupp protocol), yet median overall survival is ~14–16 months \cite{Stupp}. Light-based strategies are widely explored in GBM, for example photodynamic therapy, while preclinical monitoring commonly relies on bioluminescence imaging of luciferase-tagged glioma cells \cite{Cramer,Clark,Luwor}. However, to our knowledge there are no prior studies directly measuring endogenous ultra-weak photon emission (biophotons) from GBM cells (or astrocytes) under label-free, dark conditions; existing optical read-outs in GBM either use exogenous reporters or deliver external light. However, evidence that brain tissue emits UPE in non-tumour contexts supports biological plausibility \cite{Kobayashi,Isojima}. Establishing whether GBM and astrocytes emit detectable biophotons, and quantifying their statistical structure would therefore fill a methodological gap and may offer new insight into redox dynamics and intercellular signalling in malignant glia. \par
In this paper, we present biophoton measurements from astrocyte and glioblastoma cell samples, conducted at the Fondazione Pisana per la Scienza (FPS) using two experimental setups developed by INFN–Frascati and the University of Tor Vergata \cite{DePaolis, Benfatto1, Benfatto2}. These apparatuses, previously employed in plant germination studies \cite{DePaolis, Benfatto1, Benfatto2}, enable ultra-sensitive detection in dark conditions (2–3 counts/s) at room temperature ($\approx 25 ^{\circ}C$). For the first time in this context, Diffusion Entropy Analysis (DEA) was applied to cellular biophoton signals, allowing the identification of crucial events potentially linked to intercellular communication. The adoption of DEA marks a significant step forward in the analysis and interpretation of biophotonic data.

\section{Material and Methods}
\subsection{The experimental apparatuses}
The collaboration realized two experimental setups dedicated to the measurement of biophotons emitted by living cell cultures. The first apparatus, developed at Tor Vergata University and referred to as the "TV" setup, consists of a rectangular black PVC chamber with a cylindrical recess designed to hold the Petri dish and a rectangular lid with a photo-counting detector aligned with the center of the cylindrical recess. A small hole is located below a central circular elevation in the cylindrical recess, within the PVC chamber, slightly set back from the lateral walls. This hole allows for air exchange while minimizing light infiltration as much as possible. The second apparatus, built at the INFN National Laboratories of Frascati (LNF) and referred to as the "LNF" setup, features a cylindrical black PVC chamber with a similar cylindrical recess and a cylindrical lid housing the photo-counting detector aligned centrally. In this case, air intake is controlled through valves located on the upper part of the structure. In both the apparatuses, the detector is a H12386-210 high-speed photo-counting head (Hamamatsu Photonic Italia S.r.l, Arese (MI), Italy) \cite{Hamamatsu} powered at +5 Vcc. The phototube has a circular active area with a 5 mm radius and is extremely sensitive in the wavelength range between 230 and 700 nm with a peak sensitivity at 400 nm \cite{Hamamatsu, Hamamatsu2}.

\begin{figure}
\begin{center}
\includegraphics[width = 11 cm]{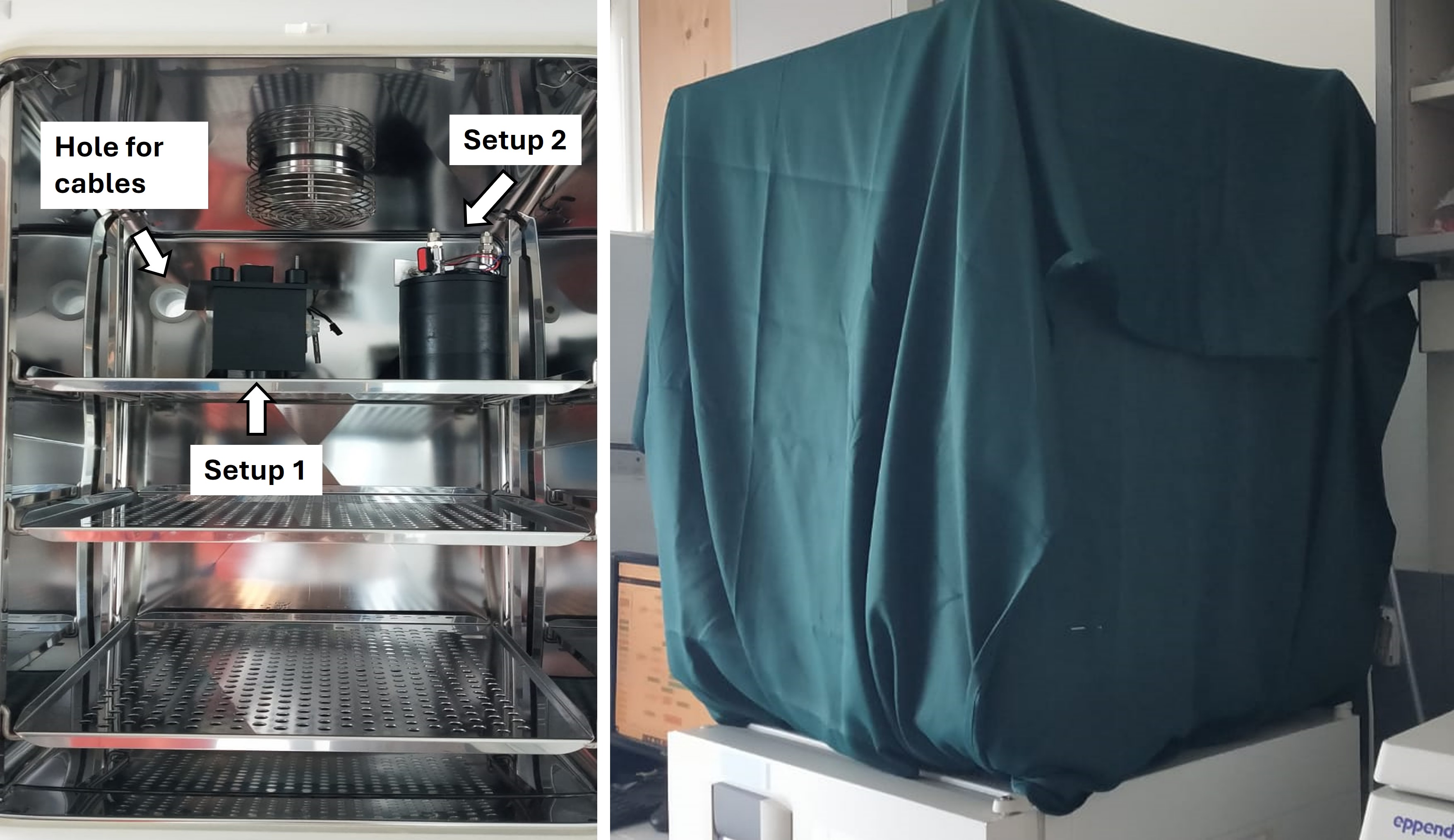}
\caption{Pictures of the final setup installed at the "Fondazione Pisana per la Scienza" (Pisa, Italy) for first measurements of biophoton emissions from cell cultures. On the left, the two setups installed inside the incubator are shown. On the right, the incubator is shielded from external light, with the two measuring devices inside. \label{fig4}}
\end{center}
\end{figure}

Both apparatuses were already used for biophoton measurements on germinating plants, delivering excellent performance and reliable results \cite{DePaolis, Benfatto2}. More details and schematic drawing of the apparatuses can be found in papers \cite{DePaolis, Benfatto2}. The setups were designed to be inserted in a dedicated incubator to preserve environmental conditions suitable for the survival and well-being of cells. Data were collected and processed using an ARDUINO board, controlled with a Node-Red-based DAQ system. The Petri dishes, with a 3 cm radius, were positioned inside the setups, exactly at the center of the detector, and at a vertical distance of 3 cm from it. This configuration was determined through a Monte Carlo simulation, which estimated the maximum geometric detection efficiency at this distance. In both setups, the center of the detector was perfectly aligned with the center of the Petri dish. The apparatuses installed in the incubator of the FPS laboratory are shown in figure \ref{fig4}.
\subsection{Cell lines and sample preparation}
Commercial human glioblastoma cell lines U87-MG, T98G, U118-MG, and commercial human brain astrocytes (HBA) were maintained at 37 $^{\circ}$C and 5$\%$ CO$_{2}$ in a humidified incubator. Cells were grown in DMEM (high glucose) supplemented with 10$\%$ fetal bovine serum (Gibco) and 1$\%$ penicillin–streptomycin (Gibco) and passaged at 70–80$\%$ confluence using 0.05$\%$ trypsin–EDTA, following the vendor’s recommendations. \par
At baseline (T0), 150,000 HBA cells and 350,000 T98G cells were seeded, respectively. After cell attachment, cultures were exposed to different experimental conditions and collected either at baseline (T0) or after 48 hours (T2). For HBA cells, one set of cultures (HBA-T0) was collected at seeding as a control. A second group (HBA-T2) was maintained inside the TOV machine for 48 hours, while additional groups were placed outside the TOV machine. Among these, some cultures were kept in the same incubator as the TOV machine, which was shielded from external light exposure, whereas others were transferred to a separate incubator that was not shielded and therefore exposed to ambient light. Similarly, T98G cells were subdivided into different groups. The control group (T98G-T0) was collected at seeding. Subsequent cultures were maintained for 48 hours (T98G-T2) either inside the TOV machine or inside the LNF machine. Additional cultures were kept outside both the TOV and LNF machines but within the same incubator, which was shielded from external light. Finally, a separate group of T98G cultures was maintained outside both machines in an unshielded incubator, directly exposed to ambient light. 
\subsection{Crystal violet staining}
At T0 or T2, dishes were rinsed in PBS, fixed in 4$\%$ paraformaldehyde (PFA) for 10–15 min at room temperature, rinsed, and stained with Crystal Violet (CV) (0.1$\%$ w/v in 20$\%$ methanol). Excess dye was removed with water, and plates were air-dried prior to imaging and quantification. Cells were de-stained using a 10$\%$ acetic acid solution, and the absorbance of the solution was then measured at 590 nm. 
\subsection{The data taking conditions}
The whole campaign of measurements was performed in the biological laboratories of Fondazione Pisana per la Scienza, where the research group of the laboratories, coordinated by Chiara Mazzanti, provided circular petri dishes with cell cultures of astrocytes and glioblastoma. The DAQ time window was fixed at 1 s \cite{Hamamatsu2}. The measurements were performed with the two apparatuses installed inside an incubator, where an external hole was drilled for the cables to exit, towards the readout chips (placed outside the incubator). The entire incubator, including the cable exit, was adequately shielded with black electrical tape, aluminium foil and a green sheet, as shown in figure \ref{fig4}. The temperature inside the incubator was 37 $^{\circ}$C to maintain optimal conditions for the life and proliferation of cells in culture. Before the measurements on cell samples, preliminary data acquisition without a target inside the machines revealed dark counts of approximately 10-20 counts per second, consistent with the data sheets provided by Hamamatsu for the detector \cite{Hamamatsu2} at the incubator temperature (37 $^{\circ}$C). This attests to the high light shielding efficiency the apparatus provides, ideal for measuring weak light signals, as expected for biophotons.
\section{Collected Data and Preliminary Observations}
\subsection{The data taking campaign}
A one-week window in September 2024 was allocated for the collaboration to acquire data, encompassing installation, measurements, and final dismantling of the setup. Given the limited timeframe, the schedule was structured as follows:

\begin{enumerate}[label=({\alph*}).]
\item About two days of measurements in complete darkness, with no samples present in either setup, aimed at assessing residual luminescence decay and background levels.
\item Two days of measurements with astrocyte cell cultures placed in both setups.
\item Two days of measurements with glioblastoma cell cultures placed in both setups.
\end{enumerate}

\begin{figure}[H]
    \centering
    \includegraphics[width=0.45\textwidth]{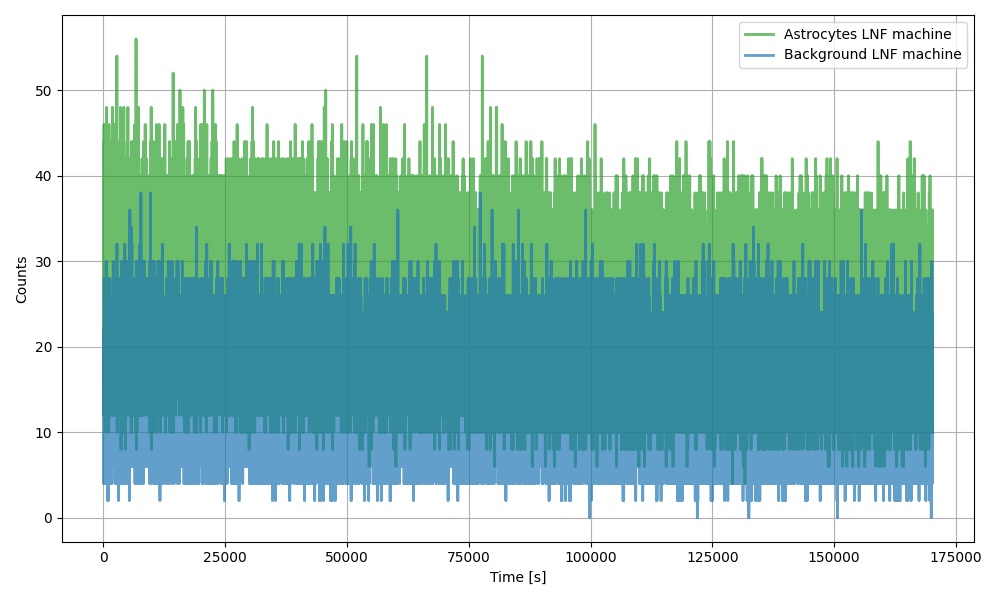}
    \hspace{0.05\textwidth}
    \includegraphics[width=0.45\textwidth]{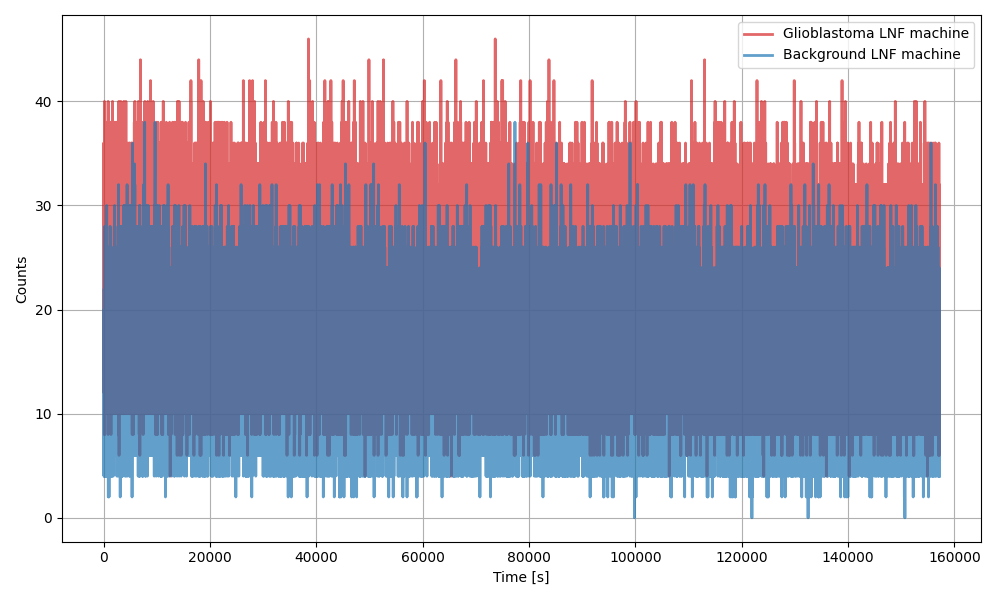}
    \caption{Comparison plots of biophotonic signals emitted by astrocyte (green) and glioblastoma (red) cell cultures concerning the background (dark counts in blue) performed with the LNF machine at the laboratories of Fondazione Pisana per la Scienza (FPS).}
    \label{fig:astr-glio_LNF}
\end{figure}

\begin{figure}[H]
    \centering
    \includegraphics[width=0.45\textwidth]{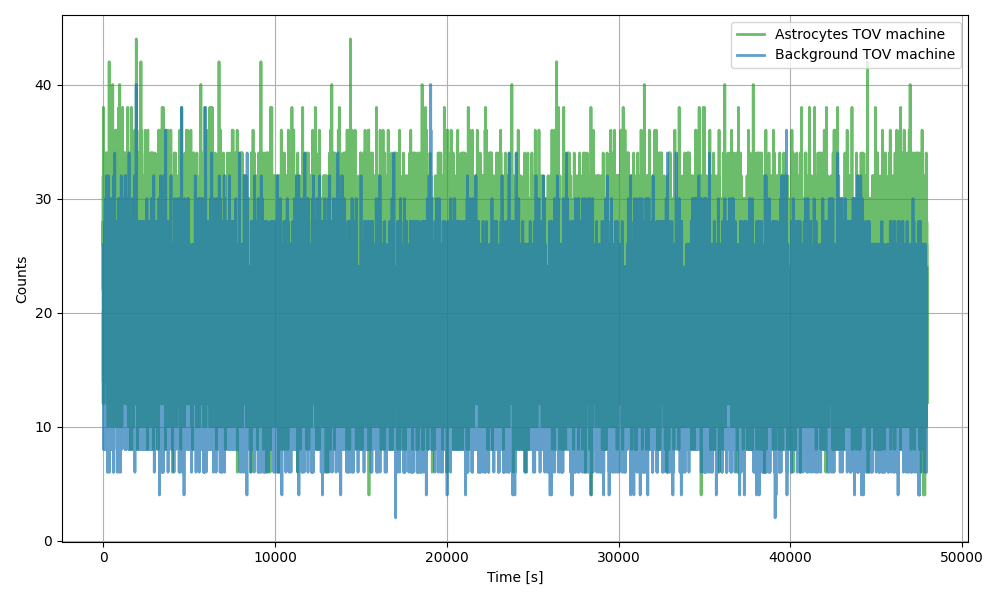}
    \hspace{0.05\textwidth}
    \includegraphics[width=0.45\textwidth]{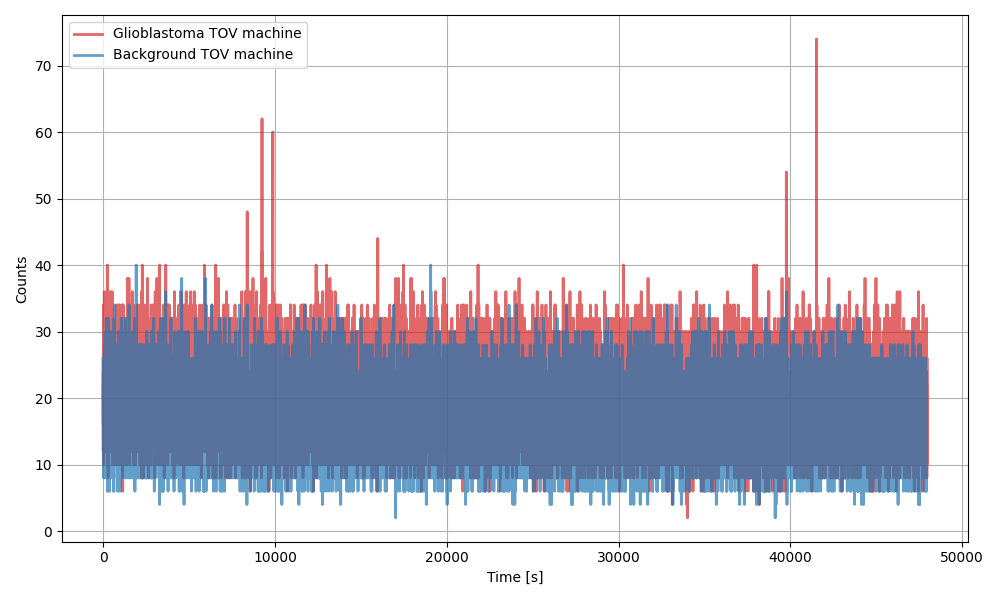}
    \caption{Comparison plots of biophotonic signals emitted by astrocyte (green) and glioblastoma (red) cell cultures concerning the background (dark counts in blue) performed with the TOV machine at the laboratories of Fondazione Pisana per la Scienza (FPS).}
    \label{fig:astr-gliob_TOV}
\end{figure}

The data collected are shown in figure \ref{fig:astr-glio_LNF} and \ref{fig:astr-gliob_TOV}. 
Looking at the graphs in figure \ref{fig:astr-glio_LNF} and \ref{fig:astr-gliob_TOV}, a clear distinction between the dark condition and the cell-containing samples was observed for both glioblastoma and astrocytes cultures measurements, attributable to the evident emission of biophotons. During the measurements, no significant increase in emission intensity was observed despite the growth in cell population. This is likely because cell proliferation does not occur uniformly, but rather locally and irregularly. Combined with the low geometrical efficiency of the setup (less than 1\%), this makes any potential increase in signal hardly detectable. Moreover, biophoton emission is not necessarily isotropic: since the cells are essentially planar, the emission may preferentially occur towards adjacent cells. Consequently, an increase in the number of cells does not necessarily correspond to a higher signal in the direction of the photocounter. Moreover, slight discrepancies between the results can be attributed to both the intrinsic differences in sample behavior—such as growth timing, cell count, and local aggregation—and the limited geometric efficiency of the detection setup.
\subsection{Statistical Analyses of the experimental data}
The signal-to-noise ratio (SNR) was estimated by calculating the signal and noise power from the recorded time series. The power of a signal, normalized to the length $N$ of the time series, was defined as:
\begin{equation}
    P_{n}=\frac{1}{N}\sum_{t=1}^{N}x(t)^{2}
\end{equation}
The SNR was then obtained as:
\begin{equation}
    SNR=\frac{P_{signal}-P_{noise}}{P_{noise}}
\end{equation}
where $P_{signal}$ and $P_{noise}$ represent, respectively, the power of signals produced during the astrocytes and glioblastoma measurements, and the power of the noise signal collected with each apparatus empty, in dark conditions and inside the incubator.\par
We estimated also the mean value of each set of data taking:
\begin{equation}
    <m>=\frac{1}{N}\sum_{t=1}^{N}x(t)
\end{equation}
and the variance:
\begin{equation}
    \sigma^{2}=\frac{1}{N}\sum_{t=1}^{N}[x(t)-<m>]^{2}
\end{equation}
 The power of signal is connected with the $\sigma$ and $<m>$ by $P_{n}=\sigma^{2}+<m>^{2}$. The variance will be fundamental for the data analysis with the Diffusion Entropy Analysis (DEA), described in the next subsections. Table \ref{tab:SNR_values} summarizes the mean values $<m>$, standard deviations ($\sigma$), skewness and estimated SNRs for the datasets acquired using both LNF and TOV setups.
\begin{table}[H]
\centering
\begin{tabular}{l l c c c c}
\toprule
Setup & Data & $<m>$ & $\sigma$ & Skewness & SNR \\
\midrule
\multirow{3}{*}{LNF} 
& Background (Dark) & 13.76 & 4.19 & 0.41 &   \\
& Astrocytes        & 22.82 & 5.51 & 0.33 & 1.68 \\
& Glioblastoma      & 20.27 & 5.06 & 0.32 & 1.09 \\
\midrule
\multirow{3}{*}{TOV} 
& Background (Dark) & 16.27 & 4.38 & 0.34 & \\
& Astrocytes        & 20.34 & 4.81 & 0.31 & 0.54 \\
& Glioblastoma      & 19.54 & 4.73 & 0.35 & 0.43 \\
\bottomrule
\end{tabular}
\caption{Table of the mean values $<m>$, standard deviations ($\sigma$), skewness and estimated Signal-to-Noise Ratios (SNRs) extracted by the analysis on dark (background), astrocytes and glioblastoma data acquired with LNF and TOV machines. The dark data were acquired with the empty setups inside the incubators.}
\label{tab:SNR_values}
\end{table}
In both setups, the dark (background) measurements show lower mean signals, as expected, since they correspond to acquisitions without biological samples. The noise levels (expressed by $\sigma$) are comparable across the different conditions and setups, with slightly higher values in the LNF configuration. When comparing the astrocytes and glioblastoma samples, we observe a progressive increase in the mean signal $<m>$ from dark to astrocytes and glioblastoma, reflecting the higher biological activity or scattering contribution of the samples. In both experimental set-ups the mean values of the counts obtained with the presence of the cellular samples are significantly higher than the same values obtained without them. To verify whether this result has a robust degree of statistical significance we performed a one-sided permutation test on the difference of means \cite{Good}. Suppose we have two independent series of observations: one corresponding to the noise ${x_{n_{1}}, ..., x_{n_{noise}}}$, and one corresponding to the signal ${y_{n_{1}}, ..., y_{n_{signal}}}$. We focus our attention on the observed difference of sample means $D_{obs}=\bar{y}-\bar{x} $ The basic idea behind this test is to assume that the signal data are not different from the noise data, and therefore the observed differences are just random fluctuations, the so-called null hypothesis H0. If so, the assignment of each observation to the “noise” or “signal” group is arbitrary, and one can pool all data into a single combined dataset: 
\begin{equation}
\mathbb{Z} = \left\{x_{n_{1}}, ..., x_{n_{noise}} , y_{n_{1}},...,y_{n_{signal}}\right\}
\end{equation}
containing $n=n_{noise}+n_{signal}$ values. From this pooled dataset, we generate surrogate datasets by randomly permuting the labels: at each iteration, we draw (without replacement) $n_{signal}$ values to form a “pseudo-signal” set and assign the remaining $n_{noise}$ values to a “pseudo-noise” set. For each such permutation we compute the mean difference:
\begin{equation}
D^{*}=\bar{y^{*}}-\bar{x^{*}}
\end{equation}
Repeating this procedure many times yields the empirical null distribution of $D^{*}$ under the assumption of no real difference between groups. The p-value is then estimated as the proportion of permutations where $D^{*} \geq D_{obs}$, with a plus-one correction to avoid zero values: 
\begin{equation}
p=\dfrac{\#\{D^{*} \geq D_{obs}\}+1}{M+1}
\end{equation}
Where M (in our case $M=20000$) is the number of performed permutation. If the null hypothesis is true, the p-value tells us how often, by shuffling the data, I get a difference as large as the one I observed. The smaller the p-value, the less likely it is that the signal series is due to a random fluctuation in the noise.  In both experimental setups we always have p-values of the order of $10^{-5}$, a very small value which indicates that the probability that the signal originates from a noise fluctuation is practically negligible. Just to have an example in case the signal was exactly the noise we would have $p \approx 0.5$. \par
From Table \ref{tab:SNR_values}, the SNR values reveal a clear difference between the two setups. These results may suggest a better sensitivity of the LNF setup, leading to more reliable detection of biological signals, especially for the glioblastoma samples. However, as already discussed in the previous section, the non-uniform distribution and proliferation of cells, varying from sample to sample, and the potentially anisotropic nature of biophoton emission are likely contributing factors to the observed discrepancies. The Table \ref{tab:SNR_values} also includes the skewness values of the count distributions within the selected time windows. Skewness quantifies the asymmetry of a distribution relative to its mean. For reference, a Gaussian distribution has a skewness of zero ($\gamma_{1}=0$) indicating perfect symmetry, while the skewness of a Poisson distribution is given by $\gamma_{1}=\frac{1}{\sqrt{\mu}}$ where $\mu$ is the parameter of the Poisson distribution. For a $\lambda = 13.76$ a Poisson distribution would exhibit a skewness of approximately $\gamma_{1}=0.269$. However, the observed values of standard deviation and skewness for the six experimental count distributions strongly deviate from these theoretical expectations. This indicates that the distributions cannot be adequately described by either Gaussian or Poisson statistics, pointing to more complex underlying dynamics in the biophotonic emission processes. \par
\begin{figure}[H]
\begin{center}
\includegraphics[width=10 cm]{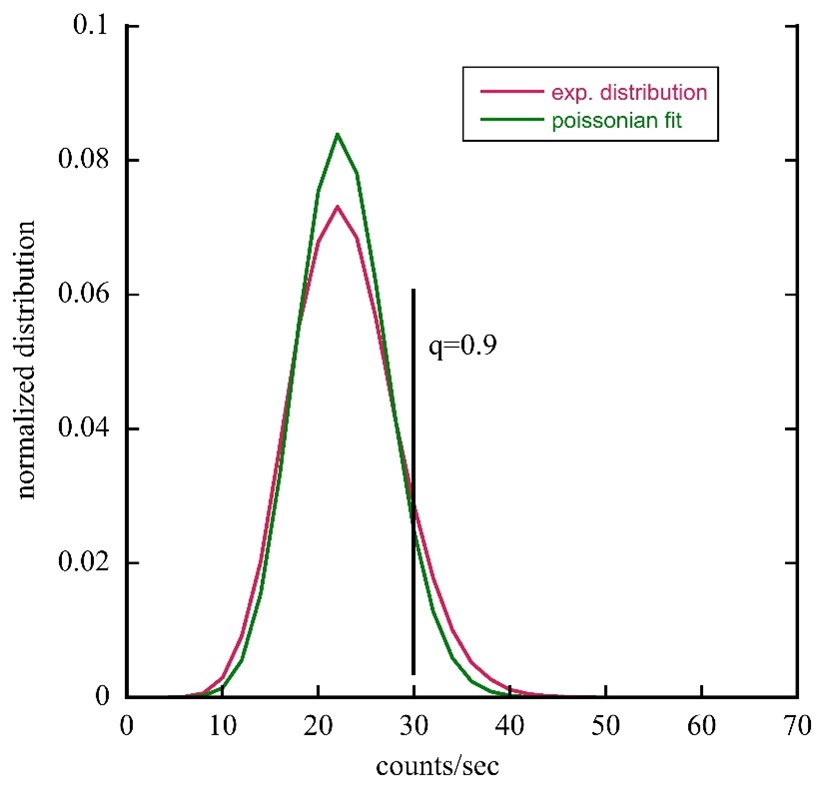}
\caption{Comparison between the count distribution function for astrocyte emission in the LNF experimental setup and the best fit performed with a Poisson function. The mean value of the best fit is $<m>=22.6$ compared to the experimental value $<m>=22.82$. The vertical black line defines the part of the distribution above the quantile $q=0.9$.\label{fignew}.}
\end{center}
\end{figure}

To go deeper into this topic, we performed an analysis of the experimental data coming from both set-up in terms of the count distribution function (CDF). Just to give an example, we show in Figure \ref{fignew} the comparison between the CDF for astrocyte emission in the LNF setup and the best fit using a Poisson function. Similar figures are obtained for all experimental data obtained in the two used setups.\par
From a visual comparison between the experimental CDFs and the corresponding fits in terms of Poisson functions, the presence of a significant tail compared to that typical of Poisson CDFs is clear. To quantify this visual observation, we used the following method. For each experimental CDF $p_{exp}(j)$ we construct the quantity $F_{exp}(k) = \sum_{j\leq k} p_{exp}(j)$ and for a given quantile q (in our case $q=0.9$) we define a threshold $thr_q = min\{k: F_{exp} (k) \geq q\}$, with this we calculate the following tail indicators, in this case the right tail:
\begin{equation}
M_{exp}(q)=\sum_{k \geq thr_q} p_{exp}(k)
\end{equation}
\begin{equation}
ES_{exp}(q)=\dfrac{\sum_{k \geq thr_q} k \cdot p_{exp}(k)}{M_{exp}(q)}
\end{equation}
The quantity $M_{exp}(q)$ represents the probability of having counts $\geq thr_q$ found with the chosen quantile, while the quantity $ES_{exp}(q)$ represents the conditional average value of the counts that exceed the threshold, i.e. how large the count is in the tail defined by the quantile. Table \ref{tab:tabnew} shows the values obtained for the two quantities defined above relating to the experimental data obtained in the two measurement systems.
\begin{table}[H]
\centering
\begin{tabular}{l c c c c}
\toprule
 & $M_{exp}(q)(LNF)$ & $ES_{exp}(q)(LNF)$ & $M_{exp}(q)(TOV)$ & $ES_{exp}(q)(TOV)$ \\
\midrule
Astr        & 0.13 & 32.29 & 0.13 & 27.86 \\
Dark & 0.10 & 21.59 & 0.14 & 23.66 \\
Glio      & 0.17 & 28.18 & 0.18 & 26.03 \\
\bottomrule
\end{tabular}
\caption{Table of the $M_{exp}(q)$ and $ES_{exp}(q)$ extracted by the analysis on background (dark), astrocytes and glioblastoma data obtained with LNF and TOV experimental setup. \label{tab:tabnew}}
\end{table}
The comparison between the two experimental configurations (LNF and TOV), which differ slightly in geometry and in the photomultiplier, although this is formally of the same type, shows that the probability of exceeding the threshold defined by the quantile is substantially consistent between the configurations. In particular, the tail probability is slightly higher in the glioblastoma samples ($\sim$0.17–0.18) compared with astrocytes ($\sim$0.13) and background ($\sim$0.10–0.14), indicating that glioblastoma cells are more frequently associated with extreme count events. More striking differences emerge in the conditional mean values. While the background measurements display the lowest average counts in the tail ($\approx$21–23), both cellular samples exhibit markedly larger values. Astrocytes in particular show the highest tail means ($\approx$28–32), clearly exceeding those of glioblastoma ($\approx$26–28) as well as background. This suggests that although extreme events occur more frequently in glioblastoma, when they occur in astrocytes they are on average more intense. Overall, these findings indicate that the presence of cells enhances both the probability and the magnitude of extreme counting events compared with background, with astrocytes characterized by rarer but stronger tails and glioblastoma by more frequent but slightly less intense tails. The small differences between LNF and TOV further suggest that the observed effects are intrinsic to the biological samples rather than to the measurement setup.


\section{Data Analysis and Results}

\subsection{The Diffusion Entropy Analysis (DEA)}
To characterize the complexity of the biophotonic signals, we applied the Diffusion Entropy Analysis (DEA) method \cite{Scafetta}. This approach is particularly suitable for studying biological systems, which often deviate from the assumptions of equilibrium statistical mechanics—such as lack of memory, short-range interactions, and non-cooperative behaviour \cite{Cakir,Grigolini}. DEA belongs to a broader class of techniques designed to quantify temporal complexity in time series data. Unlike methods based on signal compression or spectral analysis, DEA transforms the original signal into a diffusion process and measures its complexity through the growth of the Shannon entropy associated with the corresponding probability distribution. In the case of ordinary Brownian motion, this entropy grows linearly with the logarithm of time, and any deviation from this expected scaling reflects the presence of correlations or structural complexity. In our analysis, DEA revealed such deviations in the biophoton emission time series, indicating the presence of non-trivial, possibly self-organised dynamics underlying the biological activity. A more comprehensive description of the method is available in \cite{DePaolis, Benfatto1}, and a forthcoming dedicated publication will present a revised formulation of DEA specifically adapted for biophotonic signal analysis.
\subsubsection{Mathematical Formulation of DEA}
\label{subsec:math_DEA}
The Diffusion Entropy Analysis (DEA) approach starts by transforming a time series $\xi(t)$ representing, for example, the number of photons detected per second into a diffusion trajectory $x(t)$ via time integration. This trajectory reflects the cumulative behaviour of the signal and enables the investigation of its scaling properties. To study the complexity of the underlying process, we examine the probability distribution function (PDF) $p(x,l)$ of displacements over a moving window of size $l$. Then, the Shannon entropy is computed:
\begin{equation}
S(l) = -\int p(x,l)\ln p(x,l)dx
\end{equation}
If the process is characterized by a scaling law $p(x,l) \sim \frac{1}{l^\delta}F\left(\frac{x}{l^\delta}\right)$, the entropy grows as $S(l) \propto \delta \ln l$, where $\delta$ is the entropy scaling exponent and quantifies the type of diffusion: $\delta=0.5$ for normal diffusion, while $\delta \neq 0.5$ signals anomalous diffusion. However, not all anomalous scaling originates from stationary correlations. In systems with renewal events uncorrelated, memory-resetting events are separated by random waiting times $\tau$, distributed as $\Phi(\tau) \sim \tau^{-\mu}$: the source of complexity is non-stationary. These crucial events dominate the long-time behaviour when $1<\mu<3$, and standard correlation-based approaches fail to detect them. In the specific we have three cases:
\begin{itemize}
\item $1<\mu<2$: corresponds to a strongly non-stationary regime, where the waiting-time distribution has a divergent mean and variance. In this case, the processes lack a characteristic time scale, and long waiting times dominate the dynamics.
\item $2<\mu<3$: defines a weakly non-stationary regime, where the mean waiting time is finite but the variance diverges. The system exhibits subdiffusive behaviour with persistent temporal correlations and non-ergodic properties. This regime is of particular interest in the context of biophotonic signals, as it may indicate the presence of underlying temporal organization or self-structured dynamics rather than purely random activity.
\item $\mu > 3$: indicates a stationary regime, with both finite mean and variance of the waiting-time distribution. In this case, the process becomes ergodic and effectively Markovian in the long-time limit, leading to standard (Gaussian) diffusion. This behaviour is typically associated with purely random dynamics, consistent with unstructured noise lacking long-range temporal correlations.
\end{itemize}
In the conventional Diffusion Entropy Analysis (DEA) without any modification, the original time series $x(t)$ is directly integrated to generate diffusion trajectories. The scaling of the resulting entropy provides information about the overall complexity and correlations in the signal. However, this approach is primarily sensitive to Gaussian-like fluctuations and may not effectively capture rare and intermittent events, responsible for long-range memory effects. To overcome this limitation, DEA is modified using the so-called "stripes" method, which aims to highlight these crucial events explicitly. In this approach, the signal is converted into a binary sequence $z(t)$, marking the times when the original signal crosses predefined thresholds (stripes). The "stripes" were defined with a width equal to $3\sigma$, where $\sigma$ is the standard deviation of the original signal (see table \ref{tab:SNR_values}). This binarization allows for isolating and emphasizing large excursions or significant transitions in the data. A new diffusion trajectory is then constructed from the binary sequence, from which a more robust and reliable estimation of the scaling exponent $\delta$ is extracted. In fact, this modified analysis enables the detection of memory and anomalous dynamics that are otherwise hidden in the standard DEA, providing a more sensitive characterization of non-trivial temporal structures.\par
The relation between $\delta$ and the power-law index $\mu$ of the waiting-time distribution differs depending on whether stripes are used or not. Using the stripes is: $\mu=1+\frac{1}{\delta}$. Without using the stripes, $\mu$ is determined as: $\mu=4-2\delta$ \cite{Benfatto1,Benfatto2}.\par
The "stripes" approach allows distinguishing between stationary and non-stationary sources of anomalous scaling, which is crucial for understanding the complexity of the system under study. In the context of biophoton emission, the identification of weakly non-stationary dynamics, associated with $2 < \mu < 3$ and scaling exponents $\delta \ne 0.5$, suggests the presence of crucial events that punctuate the activity of the biological system. These events are not merely stochastic fluctuations, but rather reflect self-organized temporal structures that govern the dynamics over long timescales. This interpretation supports the view that living cells are complex, far-from-equilibrium systems whose activity cannot be fully described by classical statistical mechanics, but instead involves hierarchical, memory-resetting processes essential for biological function and adaptation, and may be capable of communication mechanisms among cellular groups based on biophoton emission/reception. \par
\begin{figure}
\begin{center}
\includegraphics[width=10 cm]{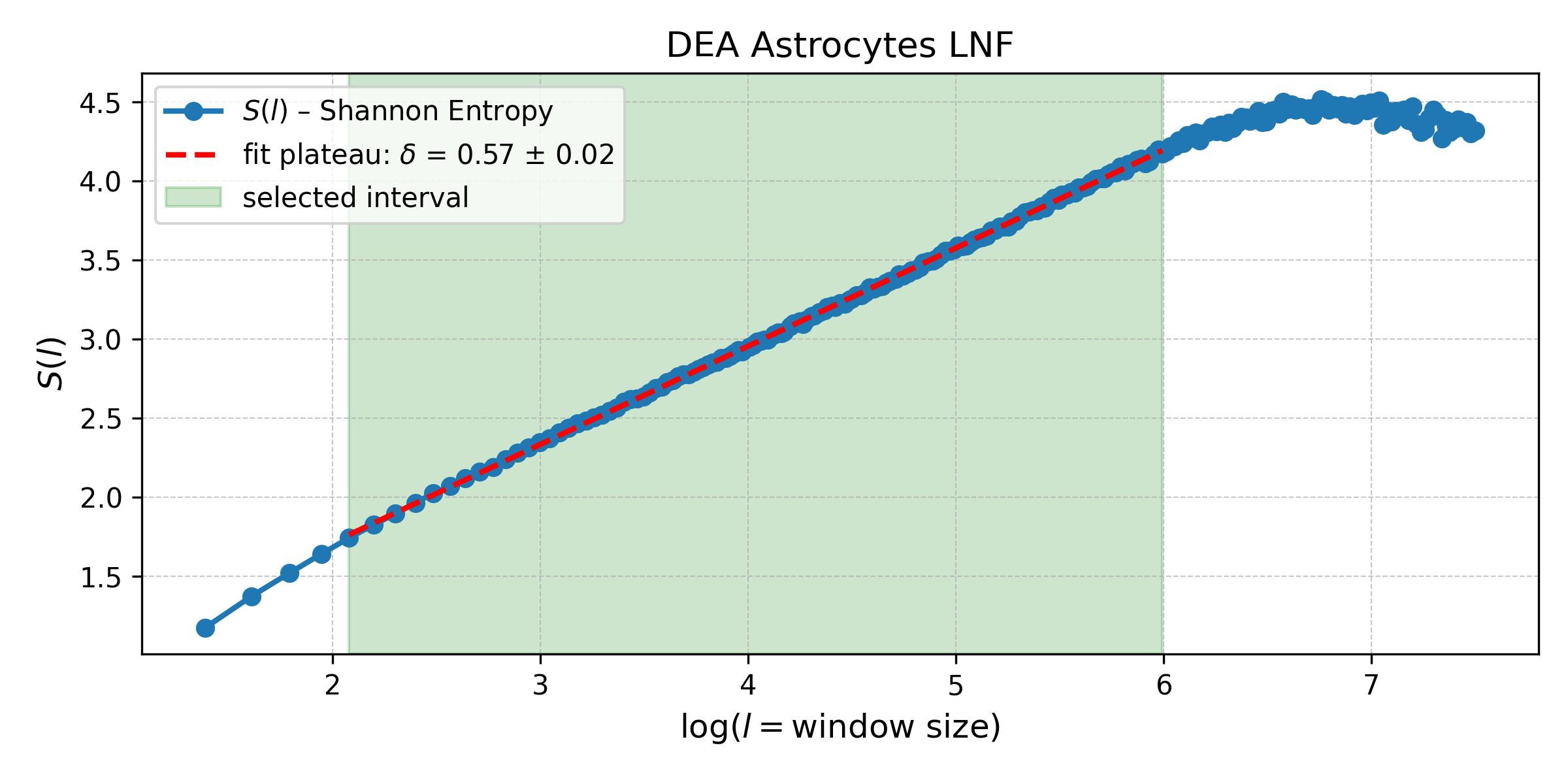}
\caption{Graph of $S(l)$ vs $\ln l$ obtained with the Diffusion Entropy Analysis (DEA) with stripes on the data collected with the LNF machine measuring biophotons emitted by astrocyte culture. As predicted by the theory, $S(l) \propto \delta \ln l$, and the linear fit in the region between $l=2$ and $l=6$ allows the extraction of the scaling factor $\delta$. For the calculation, we used a scaling window method, described in the subsection \ref{subsec:math_DEA}. The same method was applied to all data sets. Final results are reported in table \ref{tab:mu_values}. \label{DEA_fit}}
\end{center}
\end{figure}

As shown previously, the Shannon entropy is expected to follow the equation $S(l)=A + \delta \ln l$, where the scaling exponent $\delta$ is experimentally estimated through a linear fit of $S(l)$ as a function of the logarithm of $l$ (see Fig. \ref{DEA_fit} ). One of the practical challenges of DEA is the choice of the fitting interval, as statistical fluctuations strongly influence the entropy values at the boundaries. To robustly characterize the scaling behaviour of the DEA curve, we implemented a \textit{sliding window fit} procedure over the range between $ln(l)=2$ and $ln(l)=6$, where linearity is sufficiently preserved. This approach enables the local estimation of the scaling exponent $\delta$, revealing potential variations or instabilities in the scaling regime. The sliding window procedure involves the following steps:

\begin{itemize}
    \item Select a global interval $[\log(l_{min}), \log(l_{max})]$ along the horizontal axis.
    \item Define a moving window of width $w$, containing $n$ evenly spaced data points.
    \item Perform a local linear fit within each window, estimating $\delta_i$ and its uncertainty $\Delta_{\delta_i}$
\end{itemize}

The window is shifted by a step size $l$, generating a sequence of local of slope estimates:
\[
\{\delta_1, \delta_2, \delta_3, \dots, \delta_k\}.
\]
For each contiguous block of slopes of defined length (in our case we fixed this value at 5 windows), the standard deviation $(\Delta_{\delta})$ of the slopes within that block is computed. A block is classified as \emph{stable} if:
\begin{equation}
\Delta_\delta < \epsilon.
\end{equation}

In our case, $\epsilon = 0.02$. If such a plateau is found, we compute:

\begin{equation}
\delta_{plateau} = \frac{\sum_{i \in plateau} \delta_i / \Delta_{\delta_i}^2}{\sum_{i \in plateau} 1 / \Delta_{\delta_i}^2},
\end{equation}

\begin{equation}
\Delta_{plateau} = \left( \sum_{i \in plateau} \frac{1}{\Delta_{\delta_i}^2} \right)^{-1/2},
\end{equation}

i.e., the weighted mean and associated standard deviation. If no block satisfies the stability criterion, the algorithm concludes that no statistically stable plateau is present in the selected interval.
The value $\delta_{plateau}$ is then used to perform a constrained global fit of $S(l)$, fixing the slope to the plateau estimate. This allows for a direct comparison between local and global scaling behaviour, and highlights potential transitions or instabilities in the entropy dynamics. At the end, the final value of the scaling exponent $\delta$ is obtained by averaging over a restricted plateau region, where the variation of the scaling exponent is minimal.\par
For a more detailed discussion of the analysis procedures, the reader is referred to \cite{DePaolis, Benfatto1}. In addition, a technical paper is currently in preparation, presenting an updated formulation of the analysis method. This revised approach has been specifically developed to meet the demands of real-world experimental conditions and is tailored for investigating biological signals, such as biophoton emissions.
\FloatBarrier
\subsection{Results}
\subsubsection{GBM cell viability}
Figure 2 shows the results of the crystal violet assay used to assess cell viability after 48 h under the different experimental conditions. After 48 h, HBA cells showed a modest increase in viability across all conditions. Cells cultured outside the TOV machine but within the shielded incubator reached 116.9 $\pm$ 3.1$\%$, whereas those inside the TOV machine displayed 111.9 $\pm$ 6.6$\%$. HBA cells maintained in the unshielded incubator exposed to ambient light exhibited a lower increase (107.6 $\pm$ 2.6$\%$) compared with shielded conditions. \par
In contrast, T98G glioblastoma cells showed a marked proliferation under all experimental conditions. After 48 h, cells outside the machines but within the shielded incubator reached 309.6 $\pm$ 20.9$\%$, while those inside the TOV machine grew to 293.7 $\pm$ 11.8$\%$. Cultures maintained inside the LNF machine showed similar results (325.5 $\pm$ 30.7$\%$), and the highest values were observed in the unshielded incubator (324.7 $\pm$ 20.6$\%$).
These results indicate that, while astrocytes exhibited only a modest increase in cell number, glioblastoma cells proliferated robustly under all conditions, with a tendency toward higher growth in unshielded environments.
\begin{figure}[H]
\begin{center}
\includegraphics[width=10 cm]{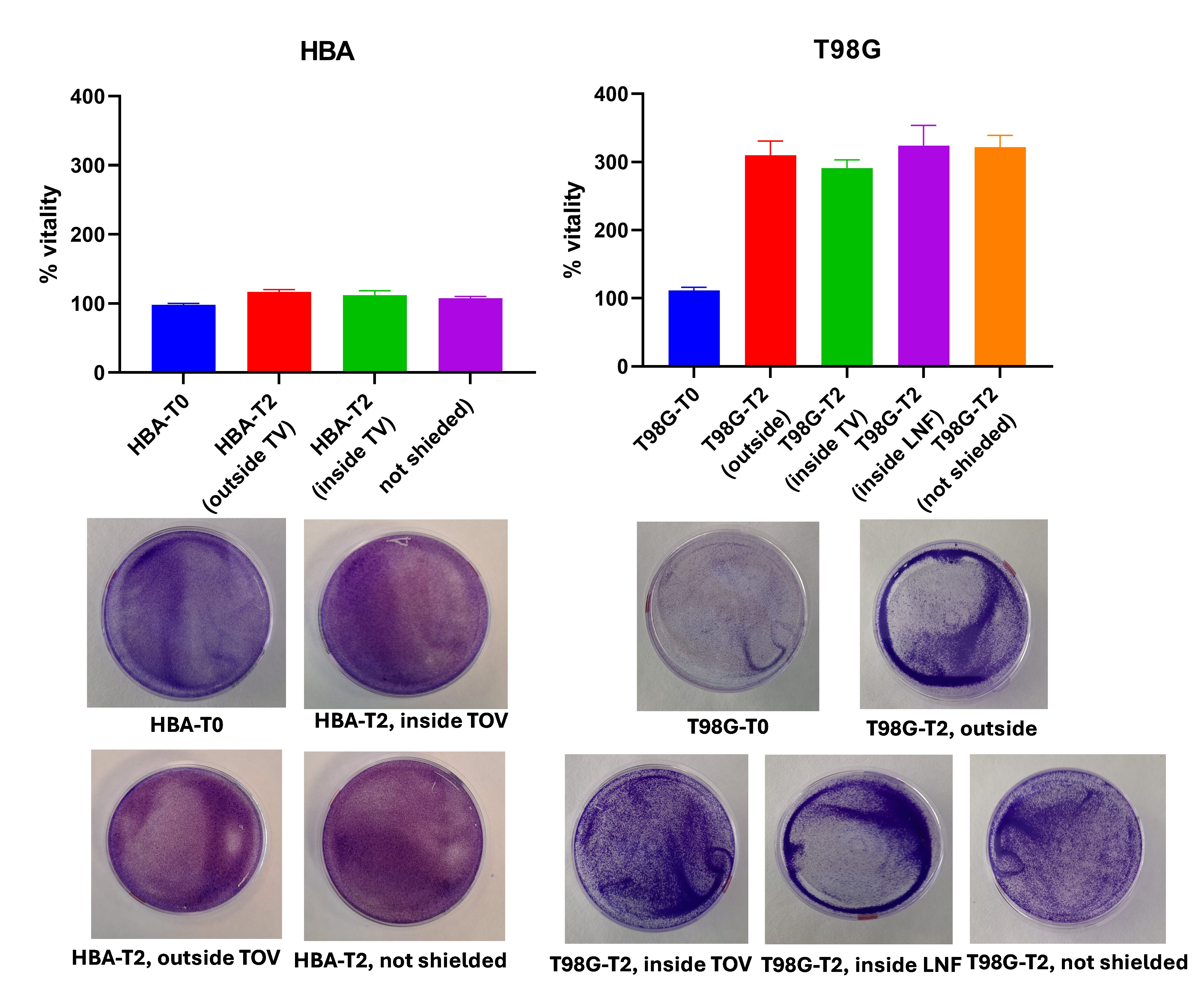}
\caption{Cell viability analysis by crystal violet assay. The histograms (top) show the percentage of viable cells quantified by crystal violet staining after 48 h under the indicated conditions. Representative images of crystal violet–stained cell cultures are shown below. Outside refers to cells maintained in the shielded incubator but outside the LNF or TOV machines; Inside TOV/LNF refers to cells cultured inside the respective machines within the shielded incubator; Not shielded indicates cultures maintained in the second incubator, which was not shielded and therefore exposed to ambient light\label{crystalviolet}.}
\end{center}
\end{figure}

\subsubsection{DEA results}
The data collected during the experimental acquisition were summarised in the figures~\ref{fig:astr-glio_LNF},~\ref{fig:astr-gliob_TOV}. Table \ref{tab:mu_values} reports the values of the scaling exponent $\mu$ obtained from DEA analyses, both without and with the application of the stripes method. As explained in previous paragraph, the comparison allows us to highlight the presence of memory effects and rare, crucial events in the biophoton emission signals. 
\begin{table}[H]
\centering
\renewcommand{\arraystretch}{2.5}
\begin{tabular}{l l c c}
\toprule
Setup & Measurement & DEA without stripes & DEA with stripes \\
\midrule
\multirow{3}{*}{LNF} 
& Background (Dark) & \makecell{$\mu$=2.76 $\pm$ 0.04 \\ $\delta$=0.62 $\pm$ 0.02} & \makecell{$\mu$=3.04 $\pm$ 0.04 \\$\delta$=0.49 $\pm$ 0.01 } \\
& Astrocytes        & \makecell{$\mu$=2.42 $\pm$ 0.02 \\ $\delta$=0.79 $\pm$ 0.01} & \makecell{$\mu$=2.75 $\pm$ 0.06 \\ $\delta$=0.57 $\pm$ 0.02} \\
& Glioblastoma       & \makecell{$\mu$=2.60 $\pm$ 0.04 \\ $\delta$=0.70 $\pm$ 0.02} & \makecell{$\mu$=2.85 $\pm$ 0.07 \\ $\delta$=0.54 $\pm$ 0.02}\\
\midrule
\multirow{3}{*}{TOV} 
& Background (Dark) & \makecell{$\mu$=2.90 $\pm$ 0.04 \\ $\delta$=0.55 $\pm$ 0.02} & \makecell{$\mu$=2.89 $\pm$ 0.07 \\ $\delta$=0.53 $\pm$ 0.02} \\
& Astrocytes        & \makecell{$\mu$=2.54 $\pm$ 0.04 \\ $\delta$=0.73 $\pm$ 0.02} & \makecell{$\mu$=2.64 $\pm$ 0.03 \\ $\delta$=0.61 $\pm$ 0.01} \\
& Glioblastoma       & \makecell{$\mu$=2.70 $\pm$ 0.02 \\ $\delta$=0.65 $\pm$ 0.01} & \makecell{$\mu$=2.79 $\pm$ 0.06 \\ $\delta$=0.56 $\pm$ 0.02} \\
\bottomrule
\end{tabular}
\caption{Values of the scaling exponent $\mu$ obtained from the Diffusion Entropy Analysis (DEA), with and without the application of the stripes method, applied to the biophoton emission signals detected from cultured astrocytes and glioblastoma cells. For the DEA with stripes, a threshold-crossing window of 3$\sigma$ counts was used to generate the binary sequence (reference to Table \ref{tab:SNR_values}).}
\label{tab:mu_values}
\end{table}
In the background (dark) measurements, both setups (LNF and TOV) show $\mu$ values closer to 3, indicating a behaviour compatible with ordinary Gaussian diffusion and absence of long-range correlations. The application of the stripes method further confirms this observation, with $\mu$ values converging exactly or even closer to 3, as expected for purely stochastic background noise. The astrocyte and glioblastoma signals exhibit distinct scaling properties, both of which differ significantly from the scaling behaviour observed in the dark signals for both systems. In all cases, the power-law exponent $\mu$ associated with the astrocyte signals is lower than that of the glioblastoma signals. Notably, both fall consistently within the range $2<\mu<3$, indicating that the time series of photocounts from the two cell types follow non-trivial statistics. This range is compatible with the presence of crucial events and fractional Brownian motion (FBM), suggesting underlying temporal correlations and complex dynamics in the biophoton emission process.\par
The scaling exponent of dark measurements obtained by applying DEA without stripes correction is slightly lower than $\mu=3$. This fact was already noted in the measurement with seeds performed by the collaboration \cite{DePaolis,Benfatto2}.  In our view, this deviation may be attributed to the presence of nonlinear components within the phototube–electronics system. By contrast, when stripes are used in the DEA analysis, the dark scaling exponent consistently approaches $\mu \approx 3$, which is the expected value for thermal-type noise.\par
The two experimental systems yield slightly different $\mu$ scaling values, with a discrepancy of about 10$\%$ in the analysis without stripe correction, which reduces to approximately 7$\%$ when stripes are applied. This difference may stem from intrinsic variations between the two experimental setups (LNF and TOV), such as differences in electronic response, phototube behaviour, or other instrumental factors. Additionally, slight observed variations likely stem from the distinct developmental profiles of the samples, combined with the inherently low geometric efficiency of the apparatus, which may amplify subtle differences in growth dynamics and spatial clustering. Despite this, the overall trend remains consistent: the scaling exponent associated with astrocyte signals is always higher than that of glioblastoma signals. In addition, differences in scaling factors were also observed even when comparing samples of the same type (astrocytes or glioblastoma) measured using different apparatuses (LNF and TOV). This variability is likely due to the non-standardized evolution of the cellular samples and the short time frame of data collection (two days). Despite starting from the same initial number of cells, the system exhibits non-deterministic behaviour, leading to differences in both cell numbers and aggregation patterns.\par
Overall, these results demonstrate that the combination of DEA with the stripes method provides a powerful tool to reveal and quantify non-trivial dynamical features and memory effects in biophoton emissions, supporting the potential of this approach to discriminate between different cell types and conditions. Furthermore, the DEA values with stripes reveal the occurrence of statistically significant crucial events, which point to coordinated, non-random activity among astrocyte and glioblastoma cells. This pattern aligns with critically-induced collective intelligence in astrocyte and glioblastoma populations, as reported in recent studies \cite{Chiara0}.

\section{Conclusions and future perspectives}
A first measurement campaign of biophotons emissions by cellular samples of astrocytes and glioblastoma was performed at the laboratories of Fondazione Pisana per la Scienza (FPS), in Pisa. The exposures lasted 2 days for each sample, and the measurements were performed with two similar setups designed and built at the National Laboratories of Frascati (LNF) and the University of Tor Vergata (TOV), in collaboration. Two days of background (dark) data were acquired for each setup installed in the incubator and empty. The data collected, shown in figures \ref{fig:astr-glio_LNF} and \ref{fig:astr-gliob_TOV}, clearly highlight a separation between the background (dark counts) and the signals recorded from astrocyte and glioblastoma samples, due to cell-derived biophoton emission. The result is supported by the signal-to-noise ratio analysis (shown in Tab. \ref{tab:SNR_values}) and confirms the high sensitivity reached by both experimental setups in measuring the weak biophotons emission in the visible range. Interestingly, the overall intensity of emission did not show a marked increase over time, despite the progressive growth of the cell population. This may be because cell proliferation tends to occur in localized and non-uniform regions, rather than evenly across the entire culture. Additionally, the low geometrical efficiency of the detection setup (less than 1$\%$) further reduces the likelihood of capturing any moderate increases in emitted signal. Finally, it is also worth noting that biophoton emission is unlikely to be isotropic: given the planar morphology of the cell cultures, emissions may be directionally biased toward neighbouring cells. As a result, a higher number of cells does not necessarily lead to a proportional increase in the signal reaching the detector, which may be an indication of the communicative nature of the emission. \par
The Diffusion Entropy Analysis (DEA) was applied to the data, both with and without the stripes method, to achieve a comprehensive characterization of ultra-weak biophoton emission from cultured astrocytes and glioblastoma cells. This approach enabled the identification of crucial events and the estimation of scaling exponents, providing insights into the underlying mechanisms governing biophoton signal generation and meaning. The DEA with no stripes results highlight the presence of anomalous diffusion dynamics and long-range memory effects in cellular biophoton emission, particularly evident in glioblastoma and astrocyte samples. The introduction of the stripes method further strengthens the evidence for the occurrence of rare and crucial events, allowing to more clearly distinguishing the non-trivial statistical features associated with living systems compared to background noise. The results obtained from both experimental setups (LNF and TOV) are quantitatively comparable, as shown in Table~\ref{tab:mu_values}, supporting the consistency and robustness of the analysis method. Moreover, the DEA reveals a clear distinction between astrocytes and glioblastoma samples, with the latter exhibiting a lower scaling exponent. A lower scaling exponent is typically associated with a shift toward more pathological or degenerative states, possibly indicating a reduced capacity for complex, self-organised activity as the system approaches cellular dysfunction or death. These analyses suggest that biophoton emission signals contain rich dynamical information beyond simple intensity measurements.  \par
In the dark condition without applying the stripes method, the estimated $\delta$ value is significantly greater than 0.5 ($\mu < 3$). This deviation likely arises from nonlinear effects associated with the measurement apparatus, as well as thermal contributions due to the system operating at 3 $^{\circ}$C. Such artefacts are effectively suppressed when using the stripes method, which acts as a filter, enhancing the detection of genuine dynamical features by reducing instrumental and thermal noise. Differences in scaling factors were also observed even when comparing samples of the same type (astrocytes or glioblastoma) measured using different apparatuses (LNF and TOV). This variability is likely due to the non-standardized evolution of the cellular samples and the short time frame of data collection (two days). Despite starting from the same initial number of cells, the system exhibits non-deterministic behaviour, leading to differences in both cell numbers and aggregation patterns.\par
The investigation of biophotons emitted by living organisms through state-of-art machines able to be highly sensitive to the signals and with the use of DEA analysis still representing, today, a new frontier for biological research, aiming to provide new insights into cellular metabolic states, reveal possible mechanisms of communication among cells through a self-organized biological signal, and potentially serve as a non-invasive diagnostic tool to differentiate between normal and tumour cells. Future studies will focus on further optimizing the detection setups with the application of a Winston cone for a better integration of the signal and of a water cooling system to maximize the performance of the counter. The main goal is to increase the geometrical efficiency and, then, the sensitivity of the apparatuses. A further optimization and deepening of the data analysis method is being performed, which will be finalized with a dedicated paper about the DEA method applied to biophoton signal detection and, more generally, to the analysis of biological signals. As future perspectives, the collaboration is working to perform measurements of neurons and other cellular cultures connected with the central nervous systems and beyond. 
Our results are consistent with a growing body of evidence indicating that biophoton emission reflects the metabolic and pathological states of cells. In vitro studies have shown that tumour cells emit significantly more biophotons compared to non-malignant cells, suggesting that such emissions could potentially serve as non-invasive diagnostic markers for malignancies (\cite{Murugan1}). Moreover, the spectral characteristics of biophoton emission, such as ratios between infrared and ultraviolet photon emissions, have been shown to effectively differentiate cancerous versus non-cancerous cell types \cite{Murugan2}. \par
In the context of glioblastoma and astrocytes, although direct measurements of biophoton emission in these specific cell types are still sparse, the marked differences observed in our crystal violet viability assay, where glioblastoma (T98G) cells exhibited much higher proliferation compared to normal astrocytes (HBA), align with the expectation that increased metabolic activity in tumour cells would correspond to elevated biophoton emission.\par
Furthermore, our application of DEA, particularly with the stripes method, has revealed distinct dynamical patterns: glioblastoma samples showed lower scaling exponents and more prominent rare, “crucial” events compared to astrocytes. These features suggest that biophoton emission signals carry rich dynamical information beyond mere intensity levels, possibly reflecting pathological disruption in self-organized cellular behaviour. The literature supports the notion of non-trivial statistical features in living systems’ biophoton emissions, often linked to complex metabolic or signalling processes \cite{Tong}.\par
Together, our findings and the existing literature suggest that biophoton emissions could serve not just as indicators of elevated metabolic or proliferative states in tumour cells, but also as encoders of underlying dynamic complexity. In glioblastoma, this may open new perspectives for non-invasive diagnostics and for exploring cellular communication mechanisms mediated through biophoton dynamics.\par
\FloatBarrier
\section*{Acknowledgements}
We gratefully acknowledge Roberto Francini, Fabio De Matteis and Alessandro Scordo for their invaluable contribution during the development and testing of the experimental setups.

\end{document}